# Well dispersed fractal aggregates as filler in polymer-silica nanocomposites: long range effects in rheology

*Nicolas Jouault [a], Perrine Vallat [a], Florent Dalmas [b], Sylvère Said [c], Jacques Jestin [a*], and François Boué [a]*

[a] Laboratoire Léon Brillouin, CEA Saclay 91191 Gif-sur-Yvette Cedex France

[b] Institut de Chimie et des Matériaux Paris-Est, CNRS UMR 7182, 2-8 rue Henri Dunant 94320 Thiais France

[c] Laboratoire Polymères, Propriétés aux Interfaces et Composites, Université de Bretagne Sud, Centre de Recherche, Rue Saint Maudé, BP 92116, 56321 Lorient cedex, France

[*] Corresponding author

Jacques.Jestin@cea.fr

**RECEIVED DATE (to be automatically inserted after your manuscript is accepted if required according to the journal that you are submitting your paper to)**

**Abstract.** We are presenting a new method of processing polystyrene-silica nanocomposites, which results in a very well-defined dispersion of small primary aggregates (assembly of 15 nanoparticles of 10 nm diameter) in the matrix. The process is based on the use of a high boiling point solvent, in which the nanoparticles are well dispersed, and a controlled evaporation procedure. The filler's fine network structure is determined over a wide range of sizes, using a combination of Small Angle Neutron Scattering (SANS) and Transmission Electronic Microscopy (TEM) experiments. The mechanical response of the nanocomposite material has been investigated both for small (ARES oscillatory shear and Dynamical Mechanical Analysis) and large deformations (uniaxial traction), as a function of the concentration of the particles in the matrix. Our findings show that with a simple tuning parameter, the silica filler volume fraction, we can investigate in the same way the structure-property correlations related to the two main reinforcement effects: the filler network contribution, and a filler-polymer matrix effect. Above a silica volume fraction threshold, we were able to highlight a divergence of the reinforcement factor, which is clearly correlated to the formation of a connected network built up from the finite-size primary aggregates, and is thus a direct illustration of the filler network contribution. For a silica volume fraction lower than this percolation threshold, we obtain a new additional elastic contribution of the material, of longer terminal time than the matrix. This cannot be attributed to the filler network effect, as the filler is well dispersed, each element separated from the next by a mean distance of 60 nm. This new result, which implies the filler-matrix contribution of the reinforcement, must include interfacial contributions. Nevertheless, it cannot be described solely with the concept of glassy layer, i.e. only as a dynamic effect, because its typical length scale extension should be much shorter, of the order of 2nm. This implies a need to reconsider the polymer-filler interaction potential, and to take into account a possible additional polymer conformational contribution due to the existence of indirect long range bridging of the filler by the chains.

I Introduction

Reinforcement of polymers, and particularly of elastomers, using inorganic particles is still an open challenge, both from the industrial and fundamental points of view. According to the characteristics of the filler (nature, shape, matrix structure, concentration) and of the elastomer matrix (interaction potential with the filler), many systems, with many associated mechanical properties can be found[1]. More recently, a specific interest has developed in the size effect, due to a better availability of much smaller particles. The first, simple idea was that the enhancement of the composite properties could be induced by increasing the specific contact area between the filler and the matrix; i.e., by reducing the size of the particle filler. This reduction could be made by downscaling to the polymer network mesh size, allowing further interaction[2,3]. The second idea was that using nanosized particles would produce a convenient extent of clustering. Controlled clustering could make it possible to control the hierarchical organization of the matrix's filler, resulting in specific filler-matrix interactions opening the way to new mechanical properties. These ideas, together with advances in characterization methods, lead to the creation of elaborate advanced composites with nanofillers, from classical carbon black[4], to silica[5,6,7,8,9] and up to carbon nanotubes[10]. Most studies used rheological approaches of various mechanical solicitations, and showed various kinds of rheological transition related to reinforcement.

In this context, two main contributions to reinforcement stand out, borrowing both from earlier research and from more recent theory. The first contribution is linked to the filler network structure: beyond the classical hydrodynamic description of Einstein[11], Guth and Smallwood[12], Batchelor[13], the idea is that in a further range of particle concentration, the particles can form a connected network (as in the percolation model), whatever their shape: simple spheres, more anisotropic rods, platelets[9,10] or complex shapes like fractals[14]. The second contribution is attributed to the interface between the matrix polymer chains and the particle filler. Beyond earlier ideas of adsorption-desorption and occluded rubber confinement, the idea that the chain dynamics could be modified in the area close to the filler object was reformulated in terms of a kind of "glassy layer", through earlier[15,16,17,18] and more recent[19] works, including NMR evidence on mobility and signature on the temperature dependence of mechanical response.

The experimental distinction between these two classes of contributions to the reinforcement mechanisms, filler network and interface, is difficult because they are often correlated:

- when starting from already complex fillers (such as carbon black or fumed silica) the structure of the filler affects the build-up, the geometry, the connectivity of the filler network, and the related interfacial interaction. The interaction potential between the chain and the filler is also associated to the filler's form factor. As discussed by Vilgis[20], the problem is multi-length and multi-time scale so that many characteristics lengths must be identified, from the nanometer range to the macroscopic one.

- when starting from individual nanoparticles, once again, an (often spontaneous) build-up of clusters brings the problem back to the former situation. However, particularly in industry, dispersion control implies a use of surfactants, grafted short linear molecules, coupling agents, which greatly and simultaneously modify – and thus correlate – the filler structure and the local interface. Trying to reduce this correlation, this same idea of controlling the filler's local dispersion, in particular the nanofillers, has been used in academic research, through more or less sophisticated chemistries like in-situ polymerization after dispersion of nanoparticles[21], or grafting polymerization on individual nanoparticles[22,23]. Other methods implied simple control of electrostatics repulsion[24], or of magnetic attraction[25], by mixing the particles with polymer beads (latex) – instead of polymer chains like in the great majority of tire processing (at least up to very recently).

In the present paper, we are presenting another strategy, which turns out to be very simple, based on a new method of nanocomposite processing, using a polystyrene matrix filled with silica nanoparticles. First, we will present the method, based on the use of a particular solvent, with a high boiling point, above the glass transition temperature $T_g$ of the polymer, and a controlled-evaporation procedure which appears to limit the attractive Van der Waals interactions effect, and thus makes it possible to obtain a particularly well defined dispersion of the silica particles in the matrix. Secondly, we will describe the fine characterization of the filler dispersion in the matrix on a very wide length range, through a combination of Small-Angle Neutron Scattering (SANS) and Transmission Electronic Microscopy (TEM). Thirdly, we will report the mechanical response of the material at low (ARES oscillatory shear plate-plate rheometer, and Dynamical Mechanical Analysis, DMA) and large deformation (uniaxial stretching). Finally, we will discuss the link between the filler structure inside the matrix at the local scale, with the mechanical behavior of the nanocomposites at the macroscopic scale.

**II Material and methods**

1 Sample preparation

The colloidal solution of silica nanoparticles is a gift from NISSAN (DMAC-ST). In this solution, the native particles, of an average radius of 5.2 nm, are electrostatically dispersed in the dimethylacetamide (DMAc), a polar solvent which is also a good solvent for the polymer used as a matrix, polystyrene (PS, Aldrich, $M_w$=280 000 g/mol, $I_p$=2, used as received). The glass transition $T_g$ of pure PS is around 100°C. A concentrated solution of PS in DMAc (10 % v/v) is mixed with a solution of silica in DMAc at various fractions, ranging from 0 to 30 % v/v. The mixtures are stirred (using a magnetic rod) for two hours. They are then poured into Teflon moulds (5 cm × 5 cm ×2.5 cm), and let cast in an oven at constant temperature $T_{cast}$ = 130 °C. This yields dry films of dimension of 5 cm × 5 cm × 0.1 cm (i.e. a volume of 2.5 cm$^3$). The specific conditions of film formulations, solvent titration, etc… are discussed in the Results section of the article. We cut some disks out of the films (diameter 1cm, thickness 1mm) for the plate-plate oscillatory shear cell, and 2 cm × 0.5 cm rectangular 1mm thick films, for DMA as well as for uniaxial stretching.

2 SANS experiments

Measurements were performed at the Laboratoire Léon Brillouin (LLB) on the SANS spectrometer called PACE. Two configurations were used: one with wavelength 17 Å, sample-to-detector distance of 4.70 m, and a collimation distance of 5.00 m, and one with wavelength 6 Å, sample-to-detector distance of 3m, and a collimation distance of 2.50 m corresponding to a total q range of 2.10$^{-3}$ Å$^{-1}$ to 0.1 Å$^{-1}$. Data processing was performed with a homemade program following standard procedures[26] with $H_2O$ as calibration standard. Small deviations, found in the spectra at the overlap of two configurations, are due to different resolution conditions and (slight) remaining contributions of inelastic, incoherent, and multiple scattering. To get the cross-section per volume in absolute units (cm$^{-1}$), the incoherent scattering cross section of H$_2$O was used as a calibration. It was estimated from a measurement of the attenuator strength, and of the direct beam with the same attenuator. The incoherent scattering background, due to protons of the matrix chains, was subtracted using a blank sample with zero silica fractions.

3 Transmission Electronic Microscopy

In order to complete on a larger scale the SANS analysis of the nanocomposite structure, conventional TEM observations were also performed on composite materials. The samples were cut at room temperature by ultramicrotomy using a Leica MZ6 Ultracup UCT microtome with a diamond knife. The cutting speed was set to 0.2mm.s$^{-1}$. The thin sections of about 40 nm thickness were floated on deionized water and collected on a 400-mesh copper grid. Transmission electron microscopy was performed on a Philips Tecnai F20 ST microscope (field-emission gun operated at 3.8 kV extraction voltage) operating at 200 kV. Precise scans of various regions of the sample were systematically done, starting at a small magnification which was then gradually increased. The slabs observed were stable under the electron beam. The sample aspect was the same in every area of every piece. Apart from a few cutting scratches, moderate buckling, very rare bubbles, and impurities, the pictures given below are completely representative of the single aspect of the sample, which thus appears homogeneous.

4 Oscillatory Shear Small Deformation Tests

Shear tests, corresponding to low deformation levels (0.5%), were carried out in the dynamic mode in strain-controlled conditions with a plate-plate cell of an ARES spectrometer (Rheometrics-TA) equipped with an air-pulsed oven. This thermal environment ensures a temperature control within 0.1 °C. The samples were placed between the two plate (diameter 1mm) fixtures at high temperature (160 °C), far above the glass transition, put under slight normal stress (around 0.5 N), and temperature was decreased progressively, while gently reducing the gap to maintain a constant low normal stress under thermal retraction. The zero gap is set by contact, the error on sample thicknesses is thus minimal; estimated at +0.010mm with respect to the indicated value. Artifacts slipping are noticeably reduced by this procedure, as checked by its reproducibility, and also by a sweeping in amplitude, at constant pulsation, which also made it possible to determine the limit of the range of linear deformation. To stay below this limit, the shear amplitude was fixed to 0.5%. Samples are stabilized at the temperature for 30 minutes before starting measurements. The reproducibility was first tested on pure PS samples with an average of 5 repetition measurements, permitting an estimatation of variations of the order of 10%. In dynamic mode, the frequency range was from 0.5 to 100 rad/s for different temperatures (from 160 to 120 °C), and time-temperature superposition was applied. The obtained multiplicative factor can be adjusted to WLF law[27] as follows

$$log(a_T) = \frac{C_1.(T_{ref} - T)}{C_2 + T - T_{ref}} \quad (1)$$

where $a_T$ is the multiplicative factor, $T_{ref}$ is the reference temperature of the master curve (in our case 143 °C), T the temperature of the measurement, $C_1$ and $C_2$ the WLF parameters. We found $C_1$=6.72 and $C_2$=98.03 °C, which is commonly obtained for PS samples[28]. For this plate and plate technique, we only focused on the samples with low silica concentration (from 0 to 5% v/v), because for higher silica fractions (> 5% v/v) the measurement reproducibility was not insured (risk of slippage).

5 Dynamic Mechanical Analysis

Rectangular pieces of film 2cm long × 0.5 cm wide were sanded down to a constant thickness of 0.8 mm. Dynamic mechanical analysis (DMA) measurements were performed on a TA DMA Q800 device in oscillatory tension mode, at fixed deformation rate (0.1%) and fixed frequency (5 Hz), at temperatures ranging from 40 °C to 300 °C with a heating rate of 10 K/min. Analysis of the oscillatory stress response is done by the software provided by TA; a preloading of 0.04 N is applied to avoid buckling. Note that 5 Hz corresponds to a short timespan, 0.2 sec. However, a measurement at 200 °C, applying a time-temperature superposition factor of 296, corresponds when using Eq. 1. to 60 sec at the reference temperature of 143 °C (equivalent to $2\pi/60 = 0.1$ rad/s). Two identical experiments (for pure PS samples) were performed to check the reproducibility of the measurements, and correct obtained values. The storage (E') and the loss (E'') moduli of the complex Young modulus (E*), and the loss factor $\tan\delta = $ (E''/E') at 5 Hz, are extracted from the measurements. The temperature at which $\tan\delta$ shows a maximum is noted $T_\alpha$.

6 Stress-strain Isotherms

Samples for stress-strain isotherms cut from films are carefully sanded down to a constant thickness ~ 1mm within a few microns. A grid of lines is drawn on the sample with a felt pen. Samples are stretched up to a pre-defined deformation value in a controlled constant-rate deformation ($d\lambda/dt = 0.005$ sec$^{-1}$, $\lambda$ being the elongation ratio L(t)/L(t=0)), at temperature T = 120 °C i.e. ~ 20 °C above the glass transition temperature of the polymer matrix ($T_g$ = 98 °C for pure PS). This corresponds to a typical time $(d\lambda/dt)^{-1}$ = 200 sec, which after applying the time-temperature superposition factor (0.0087 from 120 °C to 143 °C), corresponds at 143 °C to 1.8 sec. The tensile force $F(\lambda)$, where $\lambda = L/L_0$ defines the elongation ratio with respect to the initial length $L_0$, measured with a HBM Q11 force transducer, and converted to (real) stress inside the material $\sigma$ by dividing the force by the assumed cross-section during elongation. Indeed the deformation of the film is assumed to be homogeneous, and at constant volume; thus the cross-section decreases as $(1/\sqrt{\lambda})^2$. Apart from stress–strain curves, we analyze our data in terms of the nanocomposite reinforcement factor $R(\lambda) = \sigma(\lambda)/\sigma_{matrix}(\lambda)$, where $\sigma_{matrix}(\lambda)$ is the stress-strain curve for the polymer with a zero silica fraction.

7 Thermal Characterizations

Differential Scanning Calorimetry (DSC) measurements were performed on TA DSC Q100 under helium flow to characterize the $T_g$ of the nanocomposites. 5-10 mg of samples were put into a hermetic aluminium pan. An empty cell was used as reference. Samples were heated from 25 °C to 140 °C at a heating rate of 10 °C/min and kept at this temperature for 15 min to erase the thermal history of the materials. Then they were cooled to 25 °C at the same speed. This cycle was repeated once; reported $T_g$ was from the second heating, determined as the mid point of the heat flow step. Thermal Gravimetry Analysis (TGA) measurements were performed on TA TGA Q50 under nitrogen flow to evaluate the quantity of residual solvent and check silica volume fraction. A mass of 15 mg of the sample was heated from 25 °C to 600 °C at a heating rate of 10 °C/min and weight loss was measured. The weight fraction of residual solvent was found to be between 100 °C and 200 °C.

**III Results**

1 Composite films formulation and processing

As we said in our Introduction, we feel it is very important in nanocomposite science to obtain well-defined model systems, in which one knows how the filler is dispersed inside the matrix, to discriminate the polymer chain interfacial effects (including the dynamic ones) from effects of the network filler structure in the composites' mechanical properties. Nanocomposites of PS filled with silica nanoparticles have been extensively studied in the past decade[29,30]. The results show the difficulty of obtaining a good dispersion of the silica particles at the nanometer scale, which formed in most cases large compact aggregates (of the order of a hundred nanometers or larger) in the PS matrix. One cause of this dispersion problem may be a difficulty in controlling, during the filmification process, the evaporation speed of the solvent used (toluene, THF, methylethylketone (MEK)[29]…), and the temperature with respect to boiling point and $T_g$. Indeed various dispersions can be obtained with the same solvent[29,30]. Anyway, in our case, Dimethylacetamide (DMAc), the solvent in which silica particles are electrostatically stabilized (and which is also a good solvent for the PS), has a high boiling point, 167 °C. The casting temperature could be fixed at 130 °C, which corresponds to a relatively controlled evaporation regime, while remaining above the $T_g$ of the bulk polymer. The drying time has been determined, using the dry film reference (without silica particles), as the one for which DSC measurements showed a glass transition temperature stabilizing at $T_g$ = 98 °C , i.e. very close to the one of native polymer powder ($T_g$ = 100 °C). The residual quantity of solvent inside the composite film is found to plateau around 0.4 %wt, as determined by thermogravimetry.

2 Local structure by SANS

The structure of the silica-PS composite films has been explored by SANS. The evolution of the SANS signal as a function of the silica volume fraction $\Phi_{SiO2}$ in the films, respectively 6.6, 10.5, 15.7, 19.8 and 29.4% v/v, is presented Fig. 1. After the standard corrections and normalizations, the intensities (1/V. dΣ/dΩ, in cm$^{-1}$) are normalized by the volume fraction and corrected from the incoherent scattering of the non-filled PS matrix as follows:

$$I(Q) = I_{film} - \left((1-\phi_{SiO_2}) \times I_{matrix}\right) \qquad (2)$$

Let us now discuss the aspect of the different spectra, which evidence different Q ranges:

- at high Q, all curves superimpose perfectly after dividing by the silica fraction, indicating a good normalization of the concentration and the thickness of the samples. The scattering intensity decreases like $Q^{-4}$, which is characteristic of the scattering of a sharp well-defined interface between the native silica particles ($R_0$ = 5 nm) and the polymer matrix. At the intermediate Q range, we note an oscillation, highlighted in a $Q^3.I$ versus Q representation (see insert in Fig. 1) where it gives a maximum. Since the position of this maximum does not vary with the concentration, we can assume that it corresponds to a privileged distance inside every object, more precisely between initial spheres in contact inside the objects (in other words an internal structure factor).

**[Figure 1]**

- at low Q, the remarkable point of the whole set of data in Fig.1 is the absence of any upturn in intensity, for Q tending to 0. This indicates that if larger agglomerates existed, it would require lower Qs to detect them; in practice, TEM at low magnification (see below) shows, in the direct space, a complete absence of such agglomerates. For small concentrations (6.6 and 10.5% v/v), at low Q, the curves exhibit a plateau which is the signature of finite-size objects in the probed length scale. These objects, which we will now call "primary aggregates", are the results of the aggregation of a finite number of the native silica beads. The form factor of the primary aggregate can be approached from the most diluted scattering curve (6.6 %v/v, neglecting inter-aggregate correlations). We propose to model it by the form factor of fractal objects with the following equation[31]:

$$P_{agg}(Q) \approx N_{agg} \cdot Q^{-D_f} \frac{\int_0^\infty P_{nativebeads}(Q,R)L(R,\sigma)R^3 dR}{\int_0^\infty R^3 L(R,\sigma)dR} \quad (3)$$

where $N_{agg}$ is the number of native beads inside the primary aggregate, $D_f$ is the fractal dimension of the primary aggregates, $P_{native\,beads}(Q,R)$ is the form factor of a sphere, and $L(R, \sigma)$ is the log-normal distribution of the radius, with a variance $\sigma$). As we are dealing with the most diluted film, neglecting inter-aggregates correlations as explained above, we assume the structure factor between aggregates $S_{agg\,inter}$ to be equal to 1. The total scattering intensity can thus be expressed as:

$$I(Q) \approx \phi \Delta\rho^2 P_{agg}(Q) \times S_{agg\,intra}(Q) \quad (4)$$

where $\Delta\rho^2$ is the contrast difference between the silica and the PS matrix. The intra-aggregate structure factor, due to the repulsive interaction between initial silica beads in contact, can be expressed on the basis of the Percus-Yevick relation[32], which depends only on the radius and the concentration of the particles. We then use this structure factor $S_{agg\,intra}$ to divide the measured intensity of the film containing 6.6 % v/v of silica particles; Fig. 2 shows the result and its fit to Eq.3 for $P_{agg}(Q)$. In the insert, one can see the superposition of the intensity and the calculated structure factor before division by $S_{aggintra}$.

**[Figure 2]**

In the calculation of Eq.3, we use the radius and the log-normal polydispersity distribution ($\sigma$) of the native particles, determined from a SANS measurement of a diluted solution of particles (0.1% v/v) and fixed here to $R_0$=5nm, $\sigma$=0.36. The concentration is also fixed at the nominal value, so only the aggregation number and the fractal dimension are fitted. The result of the analysis, $N_{agg}$=15, $D_f$=2.5 (corresponding to objects of a mean radius around 30 nm), agrees well with the experimental curve. The difference between the calculation and the experimental data around 4 $10^{-2}$ Å$^{-1}$ comes mainly from the fact that we have not taken into account the polydispersity in the expression of the PY structure factor. The fractal dimension of the primary aggregates (2.5) extracted from the analysis should be also a little bit over estimated by the presence, but not clearly visible and accessible in the experimental Q range, of the inter-aggregates structure factor.

For larger silica fractions in the film, we see at low Q a second remarkable feature: a maximum appears. At first sight, the presence of this "peak" confirms the good dispersion of the particles in the polymer matrix: there is a privileged distance in the system. Let us now discuss its origin, using the evolution of the peak position as a function of the concentration (Fig. 3).

Two situations can be considered. The first situation is described by the well-known hard sphere model: the primary aggregates are distributed at random, on the condition that they cannot overlap; when increasing their concentration, they come closer to each other without being connected. This would lead to a liquid-like order showing repulsive interaction between the mass centers of the primary aggregates, and the variation of the peak position as the function of the concentration should be [33] $Q^* \sim \Phi^{(1/3)}$. The second situation is that the fractal primary aggregates connect with each other, which can be analogous to a percolation transition. In this case, Q* should correspond to the characteristic size of the network mesh, and display a $Q^* \sim \Phi^{0.88}$ scaling[34]. Plotting Q* versus $\Phi$ (Fig. 3), we obtain a slope of 0.87 +/- 0.05, suggesting that we are in this second situation for $\Phi_{SiO2} > 10\%$ (below this value no maximum is seen in our Q range). Thus SANS measurements give us a global but relevant view of both the form factor of the primary aggregates, and the organization of the filler network in the polymer matrix.

However, to correlate the rheological macroscopic properties with the local filler structure, we need to make a final check, and verify that there is no additional organization of the objects at a larger scale than the one probed with SANS. At the same time, it will also be interesting to confirm the local structure deduced from scattering, i.e. from correlations in reciprocal space, with a picture in real space. This is what we will show in the next section, using Electronic Transmission Microscopy.

**[Figure 3]**

3 Structures on a Larger Scale by TEM

Two concentrations of silica particles, 6.6 and 15.7 % v/v, have been investigated with TEM. Fig. 4 presents the results obtained at two magnifications (see bars for 200 nm and 1 μm in the top and bottom Figures respectively). The most important result is that the homogeneity of the two samples is extremely good from the 50 nm scale to the macroscopic scale.

At $\Phi_{SiO2}$ = 6.6%, the size distribution of the objects appears centered on a mean value which agrees well with the mean radius of primary aggregates deduced from the SANS analysis, i.e. 30 nm. The width of the distribution is enough to let objects of 200-300 nm at most appear. The origin of these objects is briefly discussed below. However, a further analysis of TEM would be out of this paper's scope, and we will keep using SANS results as a safe ensemble average of correlations. At $\Phi_{SiO2}$ = 15%, the real space picture confirms the model deduced from SANS. The structure keeps an "open" shape, id est does not show larger compact lumps: it shows a connected structure. We observe a characteristic net mesh size of the same order as D = 2 Π/ Q* = 2Π/ 5.3 $10^{-3}$Å$^{-1}$ ~ 120 nm, corresponding to the abscissa Q* of the repulsion peak in the SANS spectrum for 15%. Thus the important result here is that on a larger scale, up to macroscopic scale, the silica spatial distribution is homogeneous, with no connectivity signature at 5% and with connectivity at 15%.

Such changes from one picture to the next, at larger silica fractions, raise the question of the evolution of the structure with silica volume fractions. Is it related to an aggregation process specific to each concentration? Or to the progressive aggregation or percolation, according to the concentration, of initial building bricks which would be the primary aggregates, until one reaches a connected network? We will not attempt to answer this question here.

For $\Phi_{SiO2}$ = 6.6%, to evaluate the mean rim-to-rim distance between two primary clusters, we have to return to SANS data. Due to the absence of a maximum in the available range 2 to 3. $10^{-3}$ Å$^{-1}$ in the SANS curve for the diluted regime, this cannot be calculated it directly. We can make two different estimates. On the one hand, in the hypothesis of a progressive connection of primary aggregates, D is larger than the minimal distance $D_{min}$, at a larger concentration, extracted when a maximum is detectable: in figure 2, the peak appears for $\Phi_{SiO2}$ = 15%, and gives $D_{min}$ ~ 120 nm, as shown above. On the other hand, ignoring this hypothesis, we can, on a general basis, use the absence of a visible maximum at Q > 3. $10^{-3}$ Å$^{-1}$. This suggests, in a less accurate way, that this distance is larger than D = $2\pi$/ 3. $10^{-3}$ Å$^{-1}$ ~ 200 nm. Knowing the mean radius of the primary aggregates, 30nm, we can deduce that the minimal distance between the rims of two primary aggregates, for the diluted regime ($\Phi_{SiO2}$ = 6.6%), can be estimated around 60 nm in the first case, and 100 nm in the second case. This order of magnitude is also confirmed by a careful examination of a large number of real space picture (an example being figure 4(a)). Since the thickness of the slab is of the order of 50 nm here, a value of 60 nm can safely be drawn for the 3 d space.

**[Figure 4]**

4 Low deformation mechanical measurements

Low deformation measurements are interesting because the microstructure of the films shows much less alteration The first set was obtained using the shear plate-plate rheometer ARES (amplitude 0.5%). In Fig. 5(a) the variation of the elastic modulus G' is presented for a fraction of silica increasing from 0 to 5 % v/v, as a function of the product of the pulsation ω (rad/s), by the factor $a_T$ of time-temperature superposition to a reference temperature 143 °C. For further discussion, we note that the lowest value of ω.$a_T$, $10^{-2}$ rad/s, corresponds to a maximum accessible time 2 π / $10^{-2}$ rad/s = 600 sec at 143 °C. For pure PS, the curve has the usual shape in a log-log plot. At a high frequency, it shows a slope of -1/2 associated to the Rouse modes of the chains. At an intermediate pulsation ω~0.1 rad/s, a drop is observed which is characteristic of a terminal time ("creep zone", around $2\pi/\omega \sim 60$ sec, at T = 143 °C ). At a lower pulsation, a decrease with $G'(\omega) \sim \omega^2$ can be, as usual, associated to a liquid state of the polymer. The terminal time measured is in agreement with the molecular weight used here for the matrix. The height of the plateau is of order of $1.10^5$ Pa, i.e. close to the value $G^0_N$, well known for the entangled rubbery plateau of PS ($2.10^5$ Pa). The behavior of $G''(\omega)$ is given in Fig.5(b), and is also characteristic of entangled polystyrene of this molecular weight.

**[Figure 5]**

When silica is introduced, both $G'(\omega)$ and $G''(\omega)$ show clearly that, at high pulsation $\omega.a_T > 100$ rad/s, whatever the silica volume fraction, the behavior of the nanocomposite is very close to the one of the PS matrix: the Rouse modes of the chains are still observed. At intermediate and low frequencies, differences appear. At $\omega.a_T > 10$ rad/s, this is limited to a progressive increase of G' with silica fractions, analogous to elastic reinforcement as expected. But in the lowest pulsation regime, $\omega.a_T < 1$ rad/s a much more differentiated behavior is observed: adding silica greatly increases the terminal times. For the lowest silica volume fraction, it is only slightly increased compared to the one for pure PS; the creep zone stays in the same pulsation range, but a more accurate value can be extracted from the ω abscissa of the crossing of G' and G''. As soon as the silica fraction reaches 2% v/v, the terminal relaxation time become so long that the creep zone is no longer visible in the experimental window. The existence of a long elasticity time is accompanied, predictably, by a low dissipation, so that the G'' values become inferior to the G' values in the low frequency range. In such a low ω range, the slopes n, of the G' and G'' curves in log-log plot become parallel - in a rather narrow window in practice - and the ratio G'/G'', written as tan δ (ω) is equal to tan (nπ/2), as observed in a gelling process[8]. For higher silica concentrations (from 3 to 5%), this additional (with respect to pure matrix) elastic behavior is even more marked in the same low ω range. It seems at Φ>3% that the curve decay for ω tending to 0 is slower. The trend of a plateau at lower ω is more and more apparent at Φ = 4% and 5%. Comparison of Fig.5 (a) and Fig.5 (b) enable us to see the parallel evolution of G' and G''. We can note, on the curve corresponding to 4%, the same type of shoulder at the same abscissa as for pure PS, suggesting that we see a contribution from chain reptation before (i.e. at higher ω than) the final slow decay. The height at low ω of what we could call the "secondary plateau" is close to $1.10^5$ Pa. Values at intermediate pulsations have also increased, becoming of the order of $5.10^5$ up to $10^6$ Pa. This makes measurements at higher $\Phi_{SiO2}$ more delicate, due in particular to risks of slippage.

For high concentrations, we turned to DMA experiments performed at fixed frequency (5 Hz), with an amplitude deformation ratio of 0.1%. As detailed in Section II, using time–temperature superposition, measurements between 143 °C and 200°C correspond to a time range at 143°C between 0.2 sec and 60 sec. In Fig. 6, we show the evolution of the storage modulus E' and loss modulus E'' for different volume fraction of silica (from 0% v/v to 20% v/v), as a function of temperature. For pure PS (without silica), at low temperature, i.e. for $T < T_\alpha$, the curve is typical of amorphous polymer with a high elastic modulus plateau of about 2.5 GPa. It corresponds to the glassy regime. The value obtained for $T_\alpha$ is the usual one retained for the glass temperature of PS.

**[Figure 6]**

When T increases above $T_\alpha$, a steep decrease of E' is observed, followed by a shoulder corresponding to the entanglement rubbery plateau, very narrow before to be cut at slightly larger temperature by a strong decrease corresponding to the terminal relaxation. Measurements were stopped above the temperature at which the modulus becomes too weak, because the samples were flowing over during the measurement time.

In the presence of silica, different behaviors appear successively as the volume fraction increases. First, the glassy plateau keeps the same height over the whole range of $\Phi_{SiO2}$. Second, at high temperatures (T > 130 °C), three successive types of behavior can be distinguished:

- for the lowest value of $\Phi_{SiO2}$ (1% v/v), the behavior remains close to that of pure PS: the creep zone shoulder at higher temperatures keeps the same shape; it shifts slightly when passing from pure PS to 1% silica- by less than 10 °C-, after which the sample flows. Note the height of the inflexion point is of the same magnitude as the modulus obtained on the oscillatory shear plate-plate ARES rheometer (i.e. of the order of 0.1 MPa; we recall that E = 3 G in linear deformation range).

- for higher $\Phi_{SiO2}$ range, 3% v/v and 5% v/v, the most important fact is that at high T (T ~ 150 °C), the characteristic creep zone of pure PS disappears completely: the curvature is reversed, so that E' decreases slowly and seems to tend toward a plateau (~ 1MPa) at high temperatures (which however it does not reach).

- for volume fractions of 10% v/v, this plateau becomes clearly visible, and much higher ($10^7$ Pa).

- for 10 and 15 % v/v, definitely, high elastic plateaus are present, at the 5Hz frequency (10 MPa and 100 MPa)

- for 20% v/v, the plateau height has jumped to more than $10^8$ Pa, only one decade below the glassy plateau height. The slight decrease visible at the highest temperature may be an artifact. We did not measure at larger temperatures than 300 °C because this would pass the limit of matrix's thermal stability. These results will be examined further in the Discussion section.

5 High deformations: Stress-Strain Isotherms

[Figure 7]

Fig. 7(a) shows the evolution of real stress σ as a function of the elongation ratio λ for $\Phi_{SiO2}$ = 0, 1, 2, 3, 4, 5, 7.5, 10, 12.5, 15% v/v. For higher concentrations (> 15% v/v), samples did not stretch homogenously at high elongation ratios: the stress was extreme, and the samples escaped it by forming local shear bands in different zones along the sample, leading to a somehow disorganized deformation. We do not present those curves and only focus on homogenous and well-stretched samples. In these elongation measurements, all shorter processes will not produce an elastic reversible contribution, though they will contribute to the effective viscosity, but this contribution is not dominant. The set of stretching curves can be summarized by the progressive increase with $\Phi_{SiO2}$ of an initial stress jump at the beginning of the elongation, while the shape at large λ is similar, except for the shift due to the initial jump. This initial stress jump can be expressed in terms of an initial slope of stress vs. deformation ε = λ-1 in a low elongation ratio (λ<1.1) regime, giving an effective Young modulus $E_{nanocomposite}$ in this range of deformation. We have to be aware of the limitations of this analysis: the deformation ε and the initial velocity gradient dε/dt at λ close to 1 are not accurately known, so that the sample may initially be subjected to a stronger and faster deformation than the nominal values. This may, in particular, alter the filler network. However, we will see in the Discussion below that these values compare well with the DMA values at equivalent instrumental times.

For very high silica concentration ($\Phi_{SiO2}$> 10% v/v), a maximum is observed in the stress-strain curves, more and more pronounced when $\Phi_{SiO2}$ increases. Most of the time, this peak appears simultaneously with local necking, which is not discussed here. Actually, to compare with the pure matrix, a simplistic way to represent the reinforcement effect in nanocomposites is to plot the ratio of nanocomposite stress over the pure matrix stress versus λ. In Fig. 7(b) this reinforcement factor R (λ) increases notably with $\Phi_{SiO2}$ for small λ (from 1 for 1% v/v to 100 for 15% v/v). At large λ, all curves show a rapid decrease toward what seems to be a constant value. This final "value" is a more slowly increasing function of $\Phi_{SiO2}$; no strong divergence with $\Phi_{SiO2}$ is seen here. Note however that the reinforcement factor remains 4 for 15%, at an elongation ratio of 2.

6 Differential Scanning Calorimetry

Fig. 8(a) shows the Differential Scanning Calorimetry thermograms for the pure polymer and the different nanocomposites (from 1% to 20% v/v). All curves exhibit similar behavior: a slow decrease of heat flow followed by a steeper descent, provoking a step and corresponding to the glass transition temperature ($T_g$). We did not observe change of step height, i.e. of specific heat capacity $\Delta C_p$ (normalized by weight fraction of polymer) at $T_g$ which is constant as a function of silica content. The width $\Delta T$ is also constant with the increase of filler concentration. So magnitude and width of the step associated with the glass transition remain constant with the increase of silica fractions. In every case, we observe a single transition, in agreement with the spatial homogeneity of the sample.

[Figure 8]

Fig. 8(b) shows the change in such measurement of $T_g$ compared to that of pure PS for several nanocomposites (1, 3, 5, 10, 15, 20% v/v). The $T_g$ determined by DSC increases with the addition of silica. The positive shift is 6 °C for $\Phi_{SiO2}$ = 20%, i.e. of same order than for $T_\alpha$ (maximum of tan δ(ω) ) and for the maximum of E". An opposite shift (negative) has been found for other PS-Silica composites prepared in different conditions[29], but in further investigations of the same group, leading to better dispersion, this negative shift vanished[30].

**IV Discussion**

The precise description of the key parameter which governs the reinforcement mechanisms of nanocomposites depends on a thorough knowledge of the filler dispersion on the local scale. This is not always accessible, due to difficulties of sample processing (reproducibility, complexity) and suitability of the samples to characterization methods. Here, a reliable and reproducible processing technique allowed us to obtain well-defined nanocomposites, on which we were able to apply a combination of SANS and electronic microscopy to picture the nano-morphology of the silica particle inside the PS matrix. The dispersion can be summarized as follows: the system presents a silica volume fraction threshold (located around 7% v/v) between two main domains of the silica organization in the matrix. For silica volume fractions lower than this critical value, the silica particles gather in finite-sized fractal clusters called "primary aggregates", of a small mean radius ~ 30 nm, which are homogenously dispersed on a larger scale inside the matrix; they have no direct connectivity since they are separated from each other by a mean typical rim-to-rim distance of ~60 nm, which represents the minimal estimated value. For silica volume fractions larger than the critical silica volume fraction, similar primary aggregates still exist and, due to the increase of their number per volume unit, they percolate into a directly connected and continuous network. The next step is now possible: to identify the mechanical responses of the composite with the two different structural organizations, for various characteristic time ranges (at 143°C) and rates of deformation depending on the technique: shorter times (0.5-60 s) for DMA (low deformation) and (2 s) for stretching (large deformation), and longer times (200 s) for small deformation for ARES.

For the shorter times, the results are summarized in Fig. 9. The moduli $E_{nanocomposite}$ extracted from the initial slopes of the stretching curves (see Section II) have been represented in the form of the reinforcement ratio $E_{nanocomposite}$ / $E_{matrix}$ , as a function of filler concentration. We see in this plot a first linear part, then a fast divergence around 7%.

**[Figure 9]**

This behavior is confirmed on the same Fig. 9 with moduli extracted from DMA measurements at T = 1.3 $T_\alpha$, which corresponds to instrumental times at 143 °C equivalent to stretching times. The two curves show parallel behavior. Assuming a fully elastic contribution, the crossover of the reinforcement factor, which corresponds to the critical volume fraction of the connected filler network's formation, can be interpreted in terms of connectivity, more precisely of percolation between the primary aggregates. Several classical equations[11,12,13] have attempted to describe such effects in the literature, but are known to fail to reproduce the divergence. This behavior was formerly observed for the rubbery modulus obtained from DMA of polymers reinforced by fillers of higher aspect ratio (fibers, platelets[9,10]), and also formulated by Heinrich et al[14] for fractal aggregates. We have attempted to fit data to such a formula[14], but the fit is very sensitive to the value of the different fractal dimensions, and easily gives unphysical values. If we restrict ourselves to the crossing between the low $\Phi_{SiO2}$ and high $\Phi_{SiO2}$ asymptotes of the curve in Figure 8, we obtain an indicative value of $\Phi^*_{SiO2} \sim 7$ to 10%. There is thus a clear direct correlation here between such mechanical behavior in nanocomposites and the local structure of the filler in the polymer matrix: the divergence of the reinforcement factor appears at the critical silica volume fraction at which the isolated primary aggregates are closer to each other and form a connected network.

For the longer times, a second and more surprising mechanical response of the composite appears in the silica volume fraction domain, in which the filler elements are not directly connected to each other, i.e. below the silica volume fraction threshold, for $\Phi_{SiO2} < \Phi^*_{SiO2}$. Although the rims of primary aggregates are separated by a mean minimal distance of 60 nm, the oscillatory shear experiments (figure 5) and the DMA experiment at high temperature (figure 6) show an additional process corresponding to a longer terminal time, greater than the terminal time of the matrix. This means that after a decay of most of the elasticity at the matrix terminal time, an elastic fraction stays. At $\Phi_{SiO2} \geq 4\%$, the complete relaxation of this elastic fraction is beyond our accessible time range. The occurrence of such elasticity between aggregates not directly connected, but separated by very long paths through a polymer matrix of shorter terminal time, which should shortcut any additional elasticity, is, to our knowledge at least, an unexpected and original result.

The origin of this additional elastic contribution can first be discussed along the idea of the existence of a "**glassy zone**" around the particle filler. This has been proposed from earlier studies[7,15,16,17,18], up to more recent research[35] from $^1$H NMR and rheological measurements. It is suggested that this zone increases the effective volume of the filler around each particle[35], and also that it shifts the threshold for percolation between the hard regions. The EHM model[36] developed for ionomers assumes the existence around the hard zones of intermediate mobility regions of a given thickness, which merge into low mobility regions when they percolate. The EHM model has been extended to nanofillers in polymer[7]. It has also been suggested that the hard coronas act on the steric repulsion ("hard sphere model") between the filler particles, hence on the spatial correlations as observed using SANS characterization[23]. In these systems, the filler is mostly spherical particles, and volume fractions are often higher than 10%[7,35,36]. A comparison with these studies addresses the problem of the value of the thickness $\zeta$ of the glassy region around the filler. It should be equal to 30 nm in our case, considering that the lowest distance between the closest rims of two primary aggregates is 60 nm. Several of the earlier studies on $^1$H NMR proposed an estimate for $\zeta$ of a few nm (for temperatures usually above $T_g$ by several tens of °C). Similar values were proposed from the volume fraction dependence of DMA measurements in ionomers[36]. More recently, Berriot et al[35] obtained by NMR an increasing thickness with decreasing T a bit closer to $T_g$: $\zeta_{NMR}$ increasing from 2 nm at $T-T_g = 100$ °C to 3 nm for $T-T_g \sim 50$ °C. These values of $\zeta_{NMR}$ agreed with the values of effective volume with respect to nominal volume fraction extracted from mechanical reinforcement. Using a formula extracted from the thickness dependence of $T_g$ in thin films[37], the same authors could extrapolate this dependence down to lower T, in the range $T-T_g < 50$ °C, and this dependence agreed with the values of $\zeta_{mech}$ obtained from mechanical reinforcement. The latter passes 25 nm only for T very close to $T_g$ only ($\zeta_{mech} \sim 30$ nm at $T-T_g \sim 3$ °C[35]). This is much closer to $T_g$ than the temperature of our different mechanical measurements, i.e. between $T-T_g = 20$ °C and 70 °C. Since $\zeta$ $_{NMR}$ decreases fast with increasing T-T$_g$, it remains notably smaller than 25 nm and does not explain our results.

Moreover, we face another discrepancy with the measurements of Berriot et al[19,35]. In their case, the dependence of ζ over T-T$_g$ results in the failure of the usual William-Landel-Ferry-Vogel-Fulcher time-temperature superposition. As a consequence, the reinforcement factor R (T,ω) does not obey the same time temperature superposition as the matrix. In our case, conversely, the time-temperature superposition applies correctly for oscillatory shear measurements, with the same coefficients as for the pure PS matrix. If a value of 25 nm is required for ζ to explain our $\Phi_{SiO2}$= 6.6 % data, this value must be kept at 120 °C for stretching, and above for shear. This is also imposed by the DMA measurements at 5 Hz which display high reinforcements for $\Phi_{SiO2}$= 6.6 %, up to T -Tg = 70 °C. So far from T$_g$, ζ $_{NMR}$ has in most studies been estimated to a few nm. Keeping much larger values would mean a particularly strong interaction between the filler and the polystyrene. Though it is in principle possible, since the filler and the polymer are different from the ones in ref.[35] and other NMR investigations, the existence of such an interaction is in contradiction with the DSC measurements: even at = 20%, DSC gives a shift of the average T$_g$ of 6 °C only. In summary, attributing through-polymer connectivity between our aggregates to glassy zones around the nanofillers implies thicknesses larger by more than a factor 10 to commonly accepted values, and without visible temperature dependence.

Along the same lines, the origin of the additional elastic contribution could be the result of the existence of continuous glassy paths between aggregates. This has been proposed as a theory to explain shifts in T$_g$ and other dynamics in ultra-thin films[38] for which some shift in T$_g$ appears, for typical film thickness of 50 nm[39]. This analogy with thin film has also been proposed by Kumar et al.[29,30] to explain the shift in T$_g$ in some nanocomposites (which is abated in[29] but not in[30], as it can depend on the matrix polymer interaction and the processing). Note that if the orders of magnitude of the characteristic distance are the same in our system and in thin films, a marked difference appears in the sign of the shifts in T$_g$ which are negative in thin film for cases corresponding to no or weak polymer-surface interaction, while it is slightly positive in our case (figure 8). This observation implies the need to reconsider the polymer-filler interaction in our case which appears to be possibly stronger than expected. Apart from specific surface chemistry considerations, which we are not in a position to study in our system, the origin of a stronger polymer-filler interaction could be attributed to the shape of our primary aggregates; at scale between the size of the individual particles (5 nm) and the size of the aggregate (30nm) the surface is contorted. This can be associated with earlier proposals that the interaction potential between polymer and a fractal filler surface is much enhanced by its roughness[14,20]. This assumption of stronger polymer-filler interaction opens the way to an additional conformational contribution, via the bridging of the primary aggregates by some matrix chains. Saverstani et al.[40] have proposed a model and some simulations based on the formation of a mixed polymer-filler network by adsorption of the chain on the filler objects. They have showed that an additional elastic contribution appears in the modulus when the mixed network is formed for a typical rim-to-rim distance equal to twice the gyration radius of the polymer chains. For our PS chains (Mw = 280 000 g/mol, $I_p$=2,) the average $R_g$ should be ~15nm; thus a distance of 30 nm is expected, still lower than our minimal rim-to-rim distance of 60 nm, but molecular weight polydispersity could play a role in this phenomenon.

To end this Discussion, we would like to recall that secondary relaxation times were formerly observed in filled systems of well-dispersed particles[23]; in this case, a model assuming no specific lower-mobility regions, but only spatial correlations at large distances was used[41]. However, the plateau moduli obtained in this case are much lower than in the present study.

**V Summary and Conclusion**

In summary, we have been able to synthesize, over a wide range of volume fraction, polymer systems filled with very small particles (5 nm in size), aggregated at nanometer size only, and well dispersed at all larger sizes. We were then able to characterize the structure precisely over the full relevant size range, from nanometer to sample size, owing to a combination of Small Angle Neutron Scattering and Transmission Electronic Microscopy. The fillers form small primary aggregates of tens of particles, of elongated, slightly ramified shape, with a radius around 30 nm, distributed homogenously in the matrix and separated by a mean rim-to-rim distance of 60 nm. When the silica volume fraction is increased, a threshold $\Phi^*_{SiO2}$ appears, at which the primary aggregates percolate in a directly connected filler network. Two very different nanocomposite structures can thus be tuned with the simple silica volume fraction parameter. The resulting mechanical response of the material was then analyzed through different mechanical tests. Initially, with short time-spans, both stretching and DMA experiment were a simple illustration of the influence of the filler network connectivity on reinforcement properties. A direct correlation between the divergence of the reinforcement factor and the formation of the primary aggregates' connected filler network was seen. Secondly, with longer time spans (than the matrix terminal time) and small deformation (1%), a more surprising mechanical signature appeared below the silica volume fraction threshold for connectivity: the material exhibited an additional elastic contribution, with very long terminal times (not accessible). Comparing this effect with the dynamic effect attributed to a glassy layer model around the filler, the discrepancy is that the typical extension layer was around 2 nm, much lower than our rim-to-rim distance of ~ 60 nm. Similar values were however seen in $T_g$ shifts of thin polymer films (the typical thickness threshold being around 50 nm). We must also reconsider the polymer-filler interaction, which can be stronger due to the higher adsorption capacity of fractal filler compared to single spheres. In this case, the mechanical behavior could be the result of an additional conformational effect through the formation, for a typical distance of $2.R_g \sim 30$ nm, of a mixed polymer-filler network by a bridging effect on the part of the polymer chains.

**Acknowledgments**

We would like to thank the Région Bretagne and the CEA for N. Jouault's PhD Grant. The Laboratoire Léon Brillouin is a common facility of the CEA (IRAMIS) and CNRS.

**REFERENCES**

[1] J.L. Leblanc, *Prog. Polym. Sci.*, **2002**, *27*, 627-687.

[2] X. Jing, W. Zhao, L. Lan, *J. Mater. Sci. Lett.*, **2000**, *19*, 377-379.

[3] L. Flandin, T. Prasse, R. Schueler, W. Bauhoffer, K. Schulfe, J.-Y. Cavaillé, *Phys. Rev. B*, **1999**, *59*, 14349-14355.

[4] M. Klüppel, G. Heinrich, *Rubber Chem Technol*, **1995**, *68*, 623-651.

[5] A. Botti, W. Pyckout-Hintzen, D. Richter, V. Urban, E. Straube, *J. Chem. Phys.*, **2006**, 124,

[6] P. Melé, S. Marceau, D. Brown, Y. de Puydt, N. D. Albérola, *Polymer*, **2002**, *43*, 5577-5586.

[7] G. Tsagaropoulos, A. Eisenberg, *Macromolecules*, **1995**, *28*, 6067-6077.

[8] P. Cassagnau, *Polymer*, **2003**, *44*, 2455-2462.

[9] E. Chabert, M. Bornert, E. Bourgeat-Lami, J.-Y. Cavaille, R. Dendievel, C. Gauthier, J.-L. Putaux, A. Zaoui, *Mater.Sci. Eng. A.*, **2004**, *381*, 320-330.

[10] F. Dalmas, J.-Y. Cavaillé, C. Gauthier, L. Chazeau, R. Dendievel, *Composites Sciences and Technology*, **2007**, *67*, 829-839.

[11] A. Einstein, *Ann. Phys.*, **1906**, *17*, 549.

[12] (a) H.M Smallwood, *J. Appl. Phys.*, **1944**, *15*, 758,.(b) E. Guth, *J. Appl. Phys.*, **1945**, *16*, 20.

[13] G.K. Batchelor, J.T. Green, *J. Fluid Mech*, **1972**, *56*, 401- 427.

[14] G. Heinrich, M. Klüppel, T. A. Vilgis, *Current Opinion in Solid State and Materials Sciences*, **2002**, *6*, 195-203.

[15] D. Kaufman, W. P. Slichter, D.D. Davis, *J. Pol. Sci. Pol. Phys. Ed.*, **1971**, *9*, 829.

[16] J. O'Brien, E. Cashell, G.E. Wardell, V. J. McBrierty, *Macromolecules*, **1976**, *9*, 653.

[17] M. Ito, T. Nakamura, Tanaka, *J. Appl. Pol. Sci.*, **1985**, *30*, 3493 - 3504.

[18] N. K. Dutta, N. R. Choudhury, B. Haidar, A. Vidal, J. B. Donnet, L. Delmotte, J. M. Chazeau, *Polymer*, **1994**, *35*, 4293-4299.

[19] J. Berriot, H. Montes, F. Lequeux, D. Long, P. Sotta, *Macromolecules*, **2002**, *35*, 9756-9762.

[20] T. A. Vilgis, *Polymer*, **2005**, *46*, 4223-4229.

[21] J. Berriot, H. Montes, F. Martin, M. Mauger, W. Pyckhout-Hintzen, G. Meier, H. Frielinghaus, *Polymer*, **2003**, *44*, 4909-4919.

[22] G. Carrot, A. El Harrak, J. Oberdisse, J. Jestin, F. Boué, *Soft Matter*, **2006**, *2*, 1043.

[23] R. Inoubli, S. Dagréou, A. Lapp, L. Billon, J. Peyrelasse, *Langmuir*, **2006**, *22*, 6683-6689.

[24] J. Oberdisse, A. El Harrak, G. Carrot, J. Jestin and F. Boué, *Polymer*, **2005**, *46*, 6695-6705.

[25] J. Jestin, F. Cousin, I. Dubois, C. Ménager, R. Schweins, J. Oberdisse, F. Boué, *Advanced Materials*, **2008**, 20 (13), 2533-2540.

[26] P. Calmettes, *J. Phys. IV*, **1999**, *9*, 83.

[27] M. L. Williams, R. F. Landel, J. D. Ferry, *J. Amer. Chem. Soc.*, **1955**, *77*, 3701.

[28] J. D. Ferry, Viscoelastic properties of polymers, *3$^{rd}$ edition, Wiley*, **1980**, p. 641.

[29] A. Bansal, H. Yang, C. Li, K. Cho, B. C. Benicewicz, S. K. Kumar, L. S. Schadler, *Nature materials,* **2005**, *4*, 693.

[30] S. Sen, Y. Xie, A. Bansal, H. Yang, K. Cho, L. S. Schadler, S. K. Kumar, *Eur. Phys. J. Special Topics*, **2007**, *141*, 161-165.

[31] C. Rivière, F. Wilhelm, F. Cousin, V. Dupuis, F. Gazeau, R. Perzynski, *EPJE*, **2007**, *22*, 1-10.

[32] J. K. Percus, G. J. Yevick, *Phys. Rev.*, **1958**, *110*, 1.

[33] H. Benoit, D. Decker, R. Duplessix, C. Picot, P. Rempp, J.-P. Cotton, B. Farnoux, G. Jannink, R. Ober, *J. Polym. Sci.*, **1976**, *14*, 2119-2128.

[34] D. Stauffer, Taylor and Francis, London 1985.

[35] J. Berriot, H. Montes, F. Lequeux, D. Long, P. Sotta, *Eur. Phys. Lett.*, **2003**, *644*, 50-56.

[36] A. Eisenberg, J.S.Kim, Introduction to Ionomers; *Wiley-Interscience*: NewYork, 1998.

[37] J.A. Forrest, J. Mattson, *Phys. Rev. E.*, **2000**, *61*, R53.

[38] D. Long, F. Lequeux, *Eur. Phys. J. E*, **2001**, *4*, 371.

[39] M. Alcoutlabi, G. McKenna, *J. Phys. Cond. Matter*, **2005**, 17, R461-R524.

[40] A. S. Sarvestani, *Eur. Pol. J.*, **2008**, 44, 263-269. L. Yezek, W. Schaertl, C. Yongming, K. Gohr, M. Schmidt, *Macromolecules*, **2003**, *36*, 4226.

[41] S. Dagreou, B. Mendiboure, A. Allal, G. Marin, J. Lachaise, P. Marchal, L. Choplin, *J. Colloid Interface Sci.*, **2005**, *282*, 202-211.

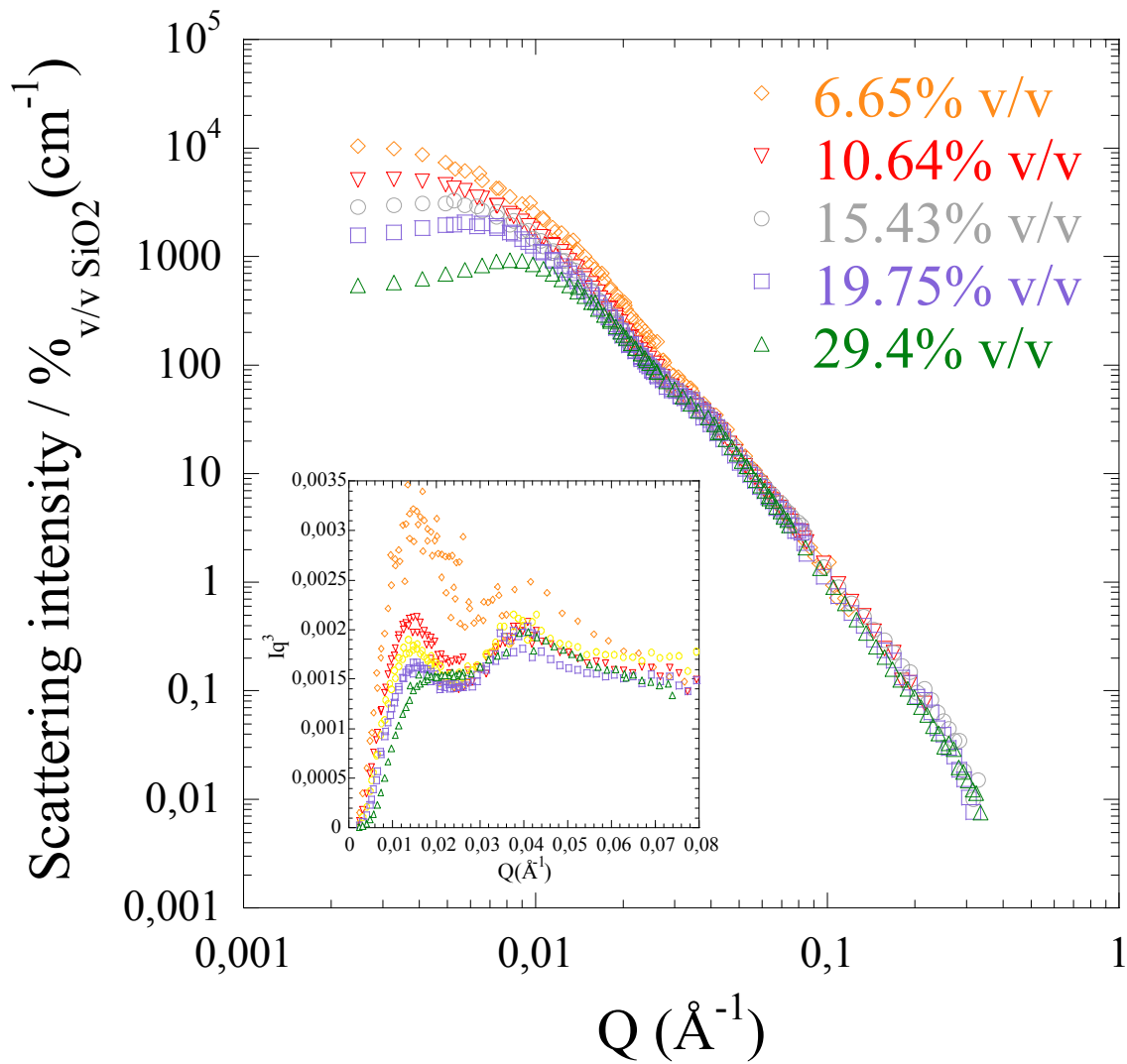

Figure 1: Scattering of the PS nanocomposite as a function of the filler concentration. The curves are normalized by the volume fraction of silica. In insert, the I.Q$^3$ versus Q representation for the same scattering curves. The oscillation, corresponding to initial spheres in contact, is highlighted around $4.10^{-2}$ Å$^{-1}$.

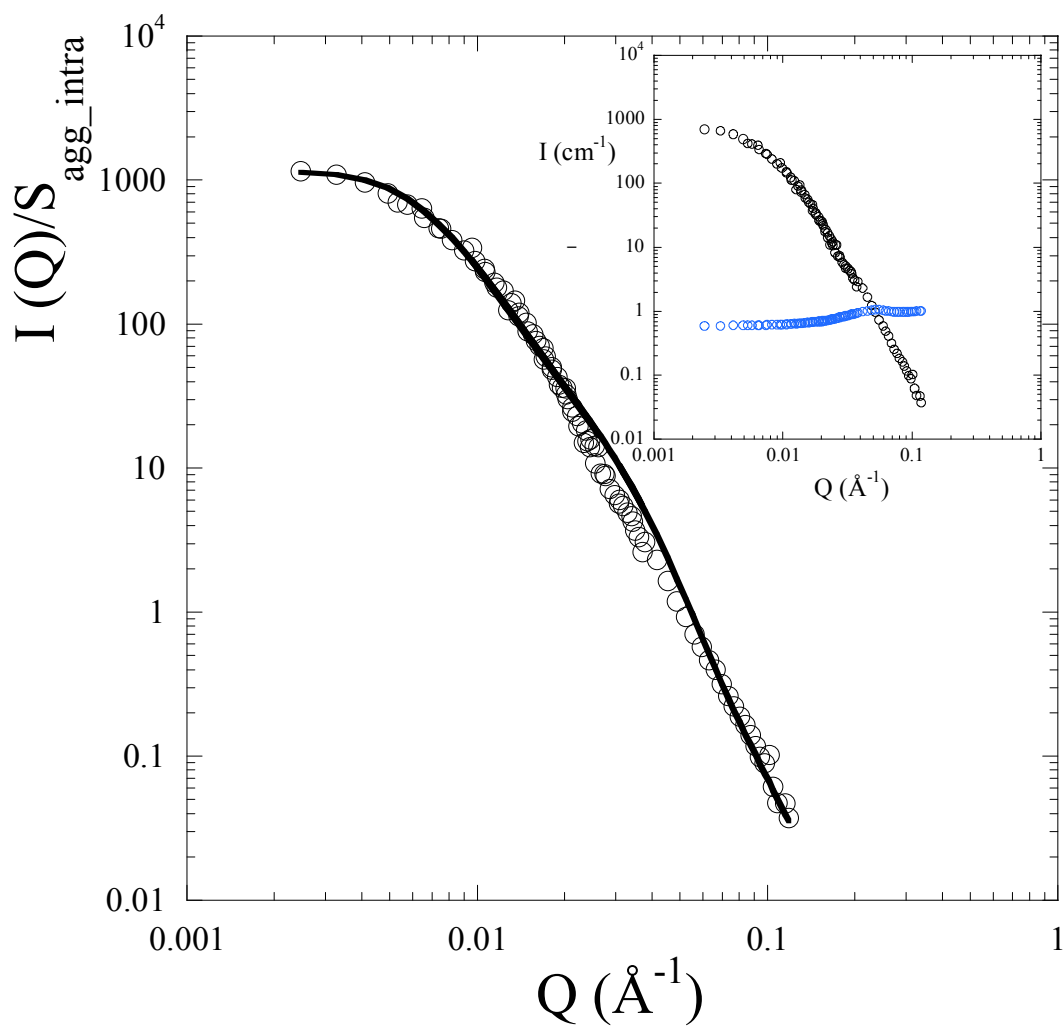

Figure 2: Form factor of the aggregates for the most diluted nanocomposite (6.6 % v/v). Black open circles are experimental data, red full line is the calculation (see in text Eq.2 and Eq.3 for details of fit). In insert, raw intensity (black open circles) and structure factor of the initial beads inside the aggregates (blue open circles).

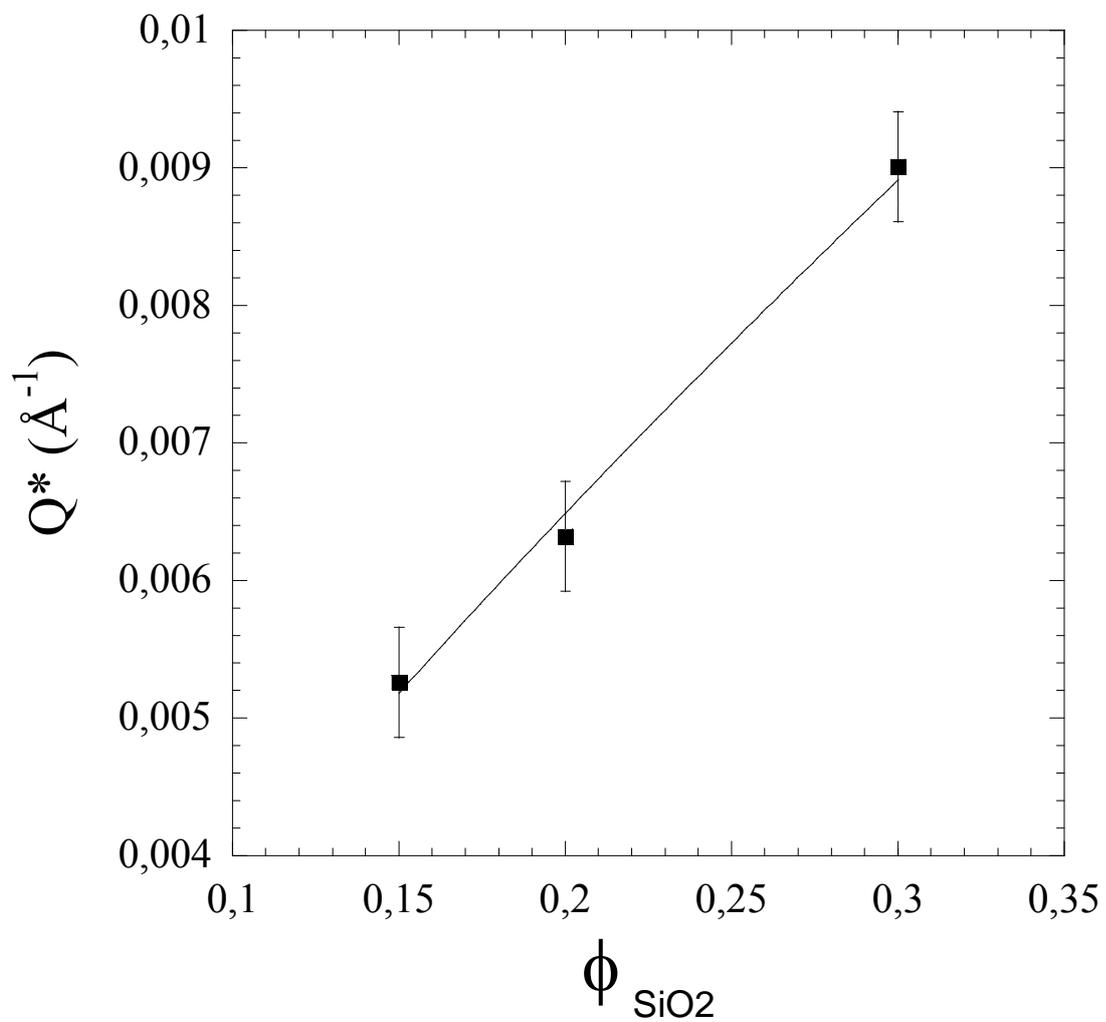

Figure 3: Variation of the filler nanocomposite concentration as a function of the scattering curve peak position (black open circle). The full line corresponds to a fit $q^* \sim \Phi_{SiO2}^{0.87}$ with error +/- 0.05 on the slope.

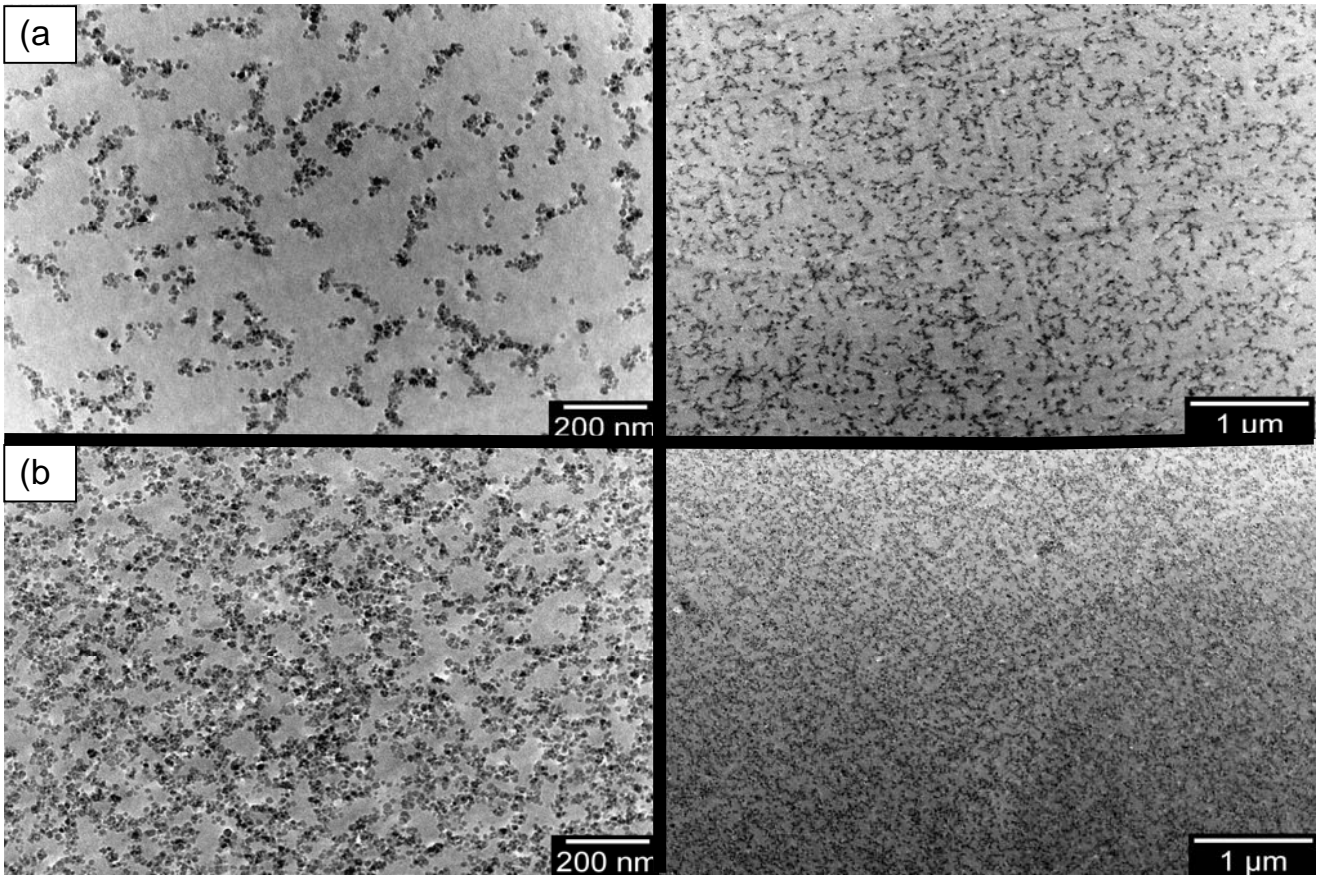

Figure 4: Transmission Electronic Microscopy on the nanocomposite filled with 6.6% v/v (a) and with 15.7% v/v (b) of silica particles. Observation at medium (above) and low (below) magnification are shown. The black zone corresponds to the silica and the grey to the polymer.

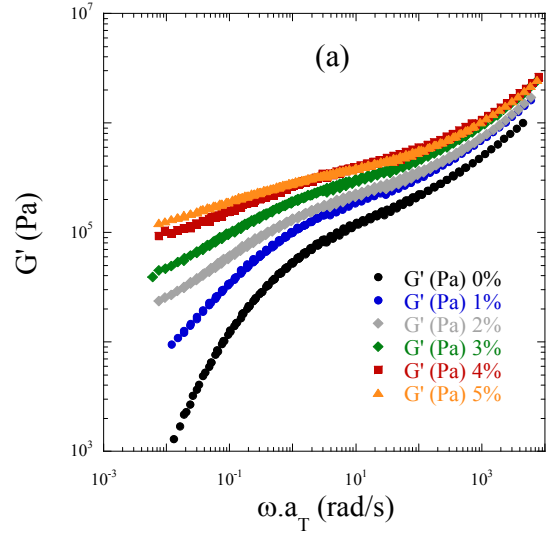

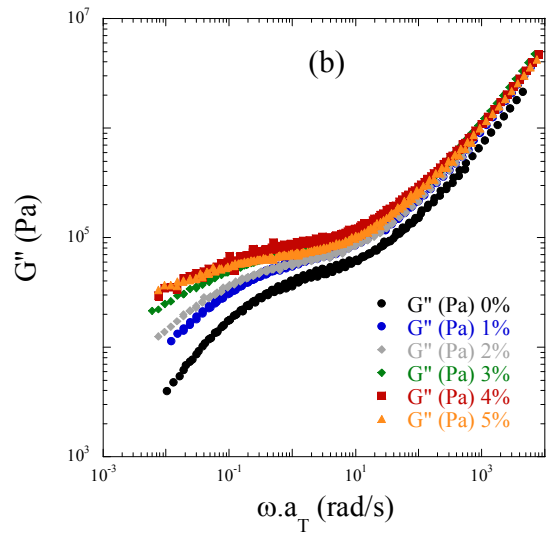

Figure 5: Elastic shear modulus G' (a) and G'' (b) as a function of pulsation ω, using time –temperature superposition ($T_0$=143°C) coefficient $a_T$ defined in the text, for different volume fractions of silica in the composite (0, 1, 2, 3, 4, and 5 % v/v).

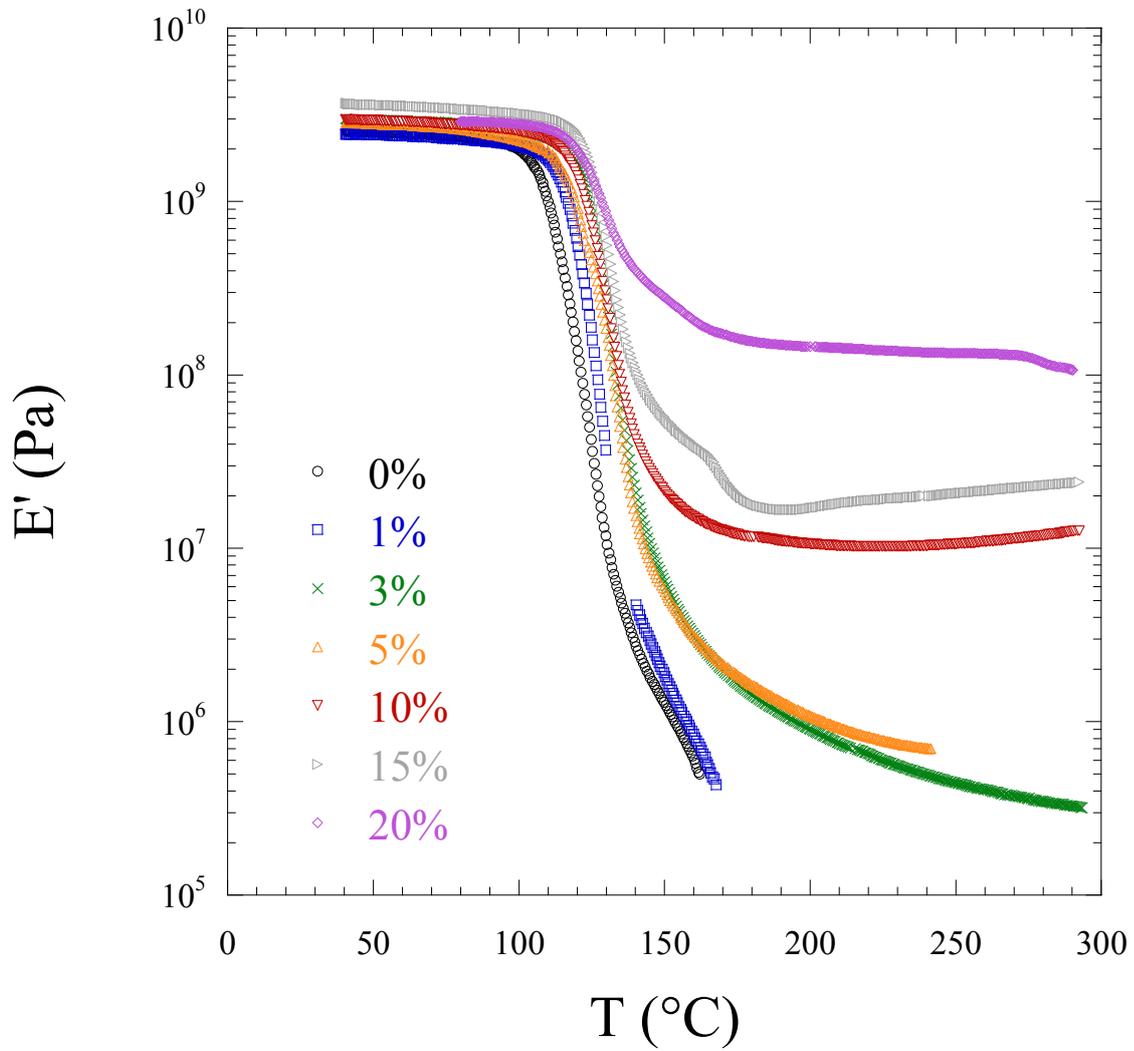

Figure 6: DMA measurements of elastic modulus E' as a function of temperature for different volume fraction of silica (from 0% to 20% v/v) (note for comparison with Fig. 5 that E'=3.G).

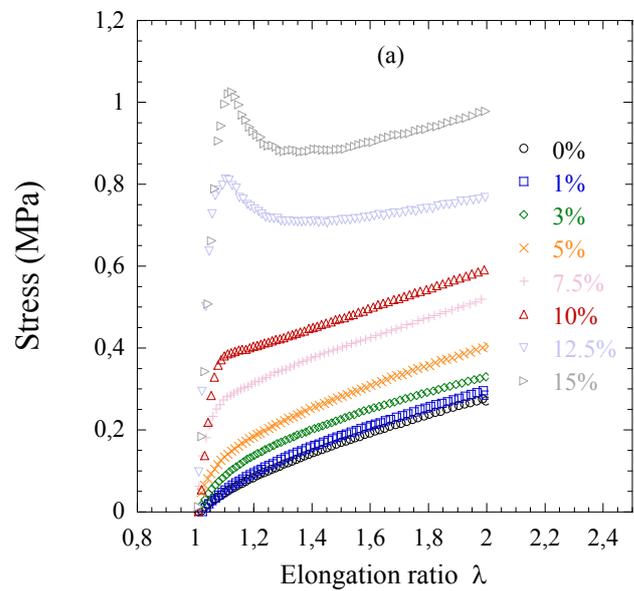

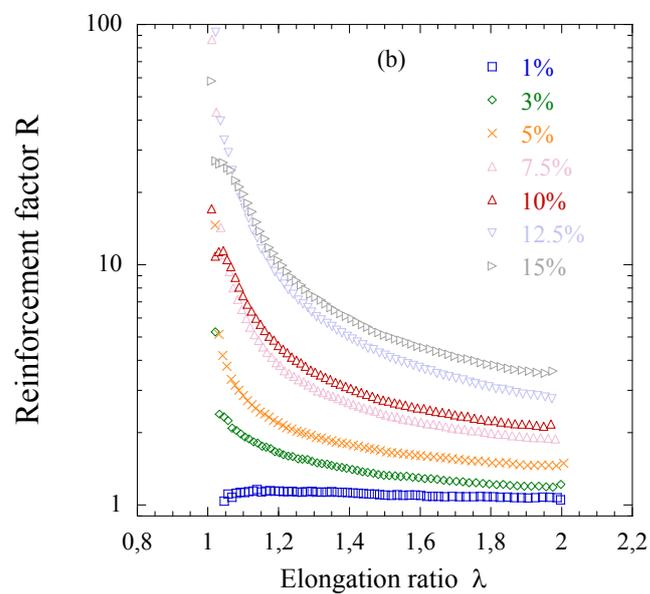

Figure 7: (a) Real stress as a function of elongation ratio for different silica volume fractions, (b) reinforcement factor as function of the elongation ratio. The experiments have been performed at $T_g+20°C$

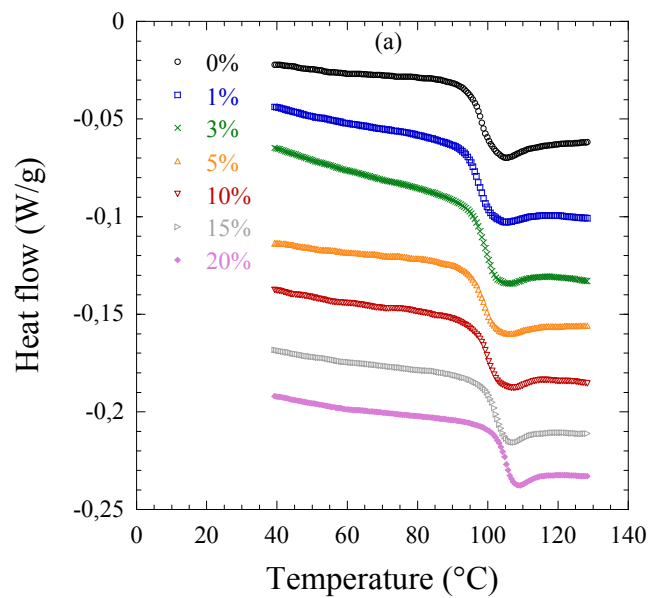

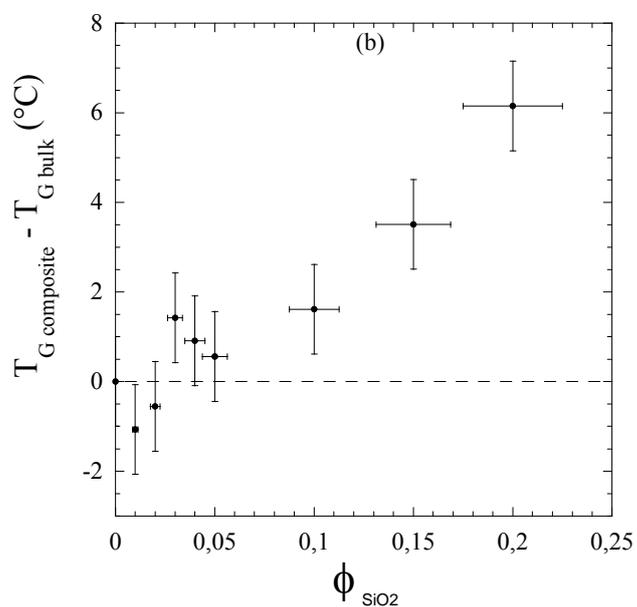

Figure 8: (a) Heat flow vs temperature for different silica content v/v %: 0, 1, 3, 5, 10, 15, 20 (To clarify, all data have been shifted along the heat flow axis). (b) $T_g$ composite − $T_g$ bulk PS as a function of silica content. Vertical errors bars are standard deviations from measurements (+/- 1°C).

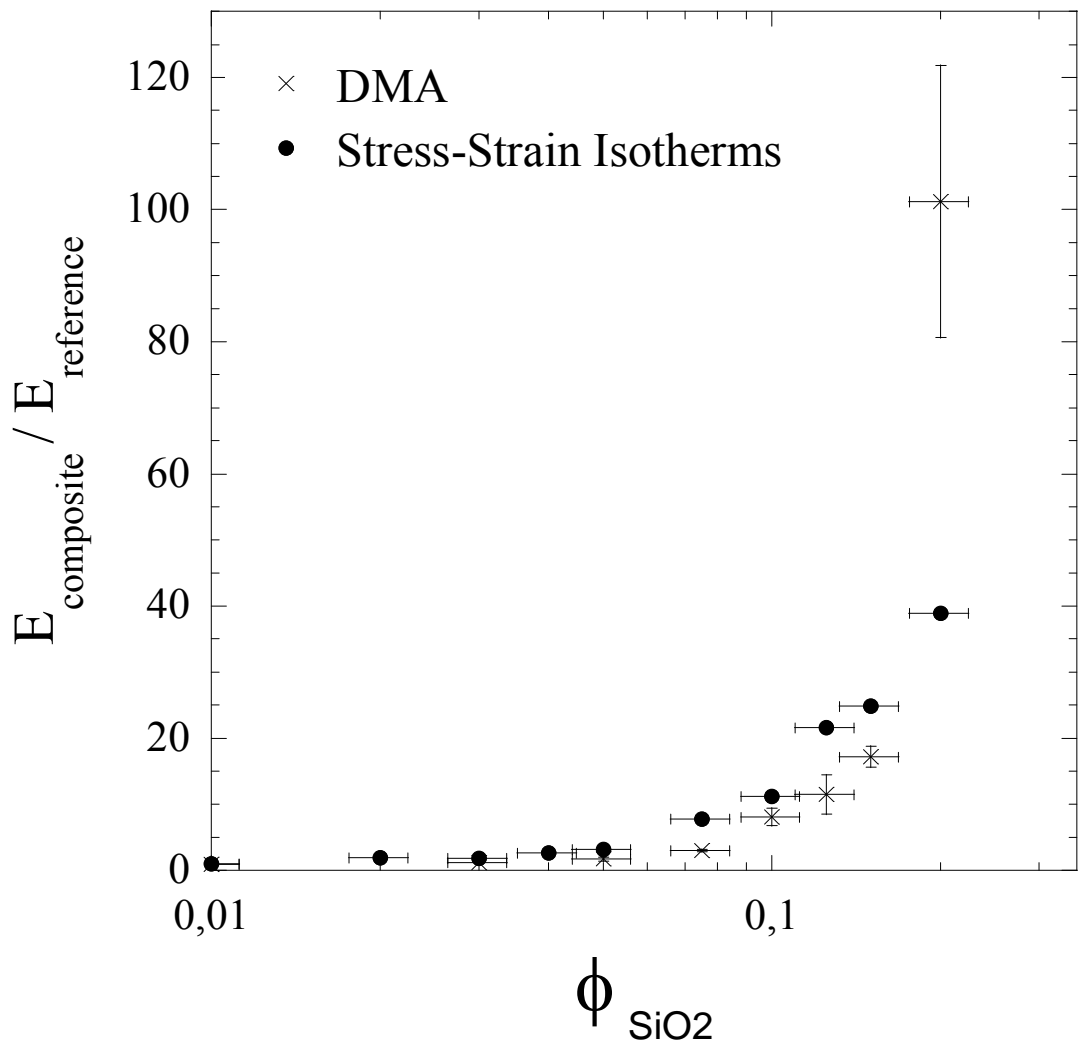

Figure 9: Reinforcement factor as a function of silica volume fraction for two different mechanical techniques: red square from DMA and blue circle from Stress-Strain isotherms

"For Table Of Contents Use Only"

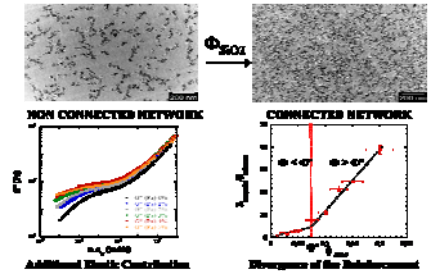